\documentclass[a4paper, 11 pt, twocolumn, final, conference]{IEEEtran}
\IEEEoverridecommandlockouts

\usepackage{graphicx}
\usepackage{subcaption}
\usepackage{times}   
\usepackage{amsmath} 
\usepackage{amssymb} 
\usepackage{mathtools}
\usepackage{algorithm}
\usepackage{amsfonts}
\usepackage{algorithm, algorithmic}
\usepackage{array}
\usepackage{multirow}
\usepackage{microtype}
\usepackage{graphics}
\usepackage{booktabs} 
\usepackage{url}
\usepackage{hyperref}
\usepackage{bm}  
\usepackage{subcaption}
\usepackage{natbib}
\usepackage{color}
\usepackage{wrapfig}
\usepackage{booktabs}
\usepackage[normalem]{ulem}
\useunder{\uline}{\ul}{}

\title{\LARGE {\bf Detecting Changes in Asset Co-Movement\\ Using the Autoencoder Reconstruction Ratio}
}

\author{Bryan Lim, Stefan Zohren, and Stephen Roberts
\thanks{B. Lim, S. Zohren and S. Roberts are with the Department of Engineering Science and the Oxford-Man Institute of Quantitative Finance, University of Oxford, Oxford, United Kingdom (email:\{blim, zohren, sjrob\}@robots.ox.ac.uk).}
}

\makeatletter

\let\c@table\c@figure
\makeatother 

\usepackage{caption}
\graphicspath{{figs/}}  

\begin{document}
\maketitle

\begin{abstract}
Detecting changes in asset co-movements is of much importance to financial practitioners, with numerous risk management benefits arising from the timely detection of breakdowns in historical correlations. In this article, we propose a real-time indicator to detect temporary increases in asset co-movements, the Autoencoder Reconstruction Ratio, which measures how well a basket of asset returns can be modelled using a lower-dimensional set of latent variables. The ARR uses a deep sparse denoising autoencoder to perform the dimensionality reduction on the returns vector, which replaces the PCA approach of the standard Absorption Ratio \citep{AbsorptionRatio}, and provides a better model for non-Gaussian returns. Through a systemic risk application on forecasting on the CRSP US Total Market Index, we show that lower ARR values coincide with higher volatility and larger drawdowns, indicating that increased asset co-movement does correspond with periods of market weakness. We also demonstrate that short-term (i.e. 5-min and 1-hour) predictors for realised volatility and market crashes can be improved by including additional ARR inputs.
\end{abstract}
\section*{Introduction}
The time-varying nature of asset correlations has long been of much interest to financial practitioners, with short-term increases in the co-movement of financial returns widely documented around periods of market stress \citep{NatureStockCorrelationsUnderStress,RiskFactorBasedCorrelation,IncreasedCorrelationInBearMarkets,AsymmetricCorrelationDynamics,CorrelationSpikesInBearMarketsMIT}.  From a portfolio construction perspective, short-term increases in asset correlations have frequently been linked to spikes in market volatility \citep{IncreasedCorrelationInBearMarkets} -- with explanatory factors ranging from impactful news announcements \citep{NewsAnnouncementReactions} to ``risk-on risk-off'' effects \citep{RiskOnOff} -- indicating that diversification can breakdown precisely when it is needed the most. In terms of systemic risk, increased co-movement between market returns is viewed as a sign of market fragility \citep{AbsorptionRatio}, as the tight coupling between markets can deepen drawdowns when they occur, resulting in financial contagion. The development of methods to detect real-time changes in returns co-movement can thus lead to improvements in multiple risk management applications.

Common approaches for co-movement detection can be broadly divided into two categories. Firstly, multivariate Gaussian models have been proposed to directly model time-varying correlation dynamics -- with temporal effects either modelled explicitly \citep{NewsAnnouncementReactions,RiskOnOff}, or incorporated simply by recalibrating covariance matrices daily using a sliding window of data \citep{NatureStockCorrelationsUnderStress}. More adaptive correlation estimates can also be provided by calibrating models using high-frequency data -- as seen in the multivariate HEAVY models of \cite{MultivariateHEAVY}. However, several challenges remain with the modelling approach. For instance, na\"ive model specification can lead to spurious statistical effects  \citep{SupriousCorrelationSpikes} reflecting correlation increases even when underlying model parameters remain unchanged. Furthermore, given that correlations are modelled in a pair-wise fashion between assets, condensing a high-dimensional correlation matrix into a single number measuring co-movement changes can be challenging in practice.     

An alternative approach adopts the use of statistical factors models to capture common movements across the entire portfolio, typically projecting returns on low-dimensional set of latent factors using principal component analysis (PCA) \citep{AbsorptionRatio,CorrelationSpikesInBearMarketsMIT,CrossCorrelationsAsSystemicRiskIndicator}. The popular Absorption Ratio (AR), for example, is defined by the fraction of the total variance of assets explained or absorbed by a finite set of eigenvectors \citep{AbsorptionRatio} -- which bears similarity to metrics to evaluate eigenvalue significance in other domains (e.g. the Fractional Spectral Radius of \cite{FSR}). Despite their ability to quantify co-movement changes with a single metric, with a larger AR corresponding to increased co-movement, PCA-based indicators require a covariance matrix to be estimated over a historical lookback window. This approach can be data-intensive for applications with many assets, as a long estimation window is required to ensure  non-singularity of the covariance matrix \citep{CorrelationSpikesInBearMarketsMIT}. Moreover, non-Gaussian financial returns \citep{NonGaussian} can violate the normality assumptions required by PCA, potentially making classical PCA unsuitable for high-frequency data. While alternatives such as Independent Component Analysis (ICA) have been considered for modelling non-Gaussian financial time series data \citep{ICA}, they typically maintain the assumption that assets are linear combinations of driving factors -- and potential improvements can be made using non-linear approaches. As such, more adaptive non-parametric methods are hence needed to develop real-time indicators for changes in co-movement.

Advances in deep learning have demonstrated the benefits of autoencoders architectures for dimensionality reduction in complex datasets \citep{DeepLearningBook}. In particular, autoencoders have had notable successes in feature extraction in images \citep{AutoencoderImageExample,AutoencoderCT,AutoencodersMedicalExample}, vastly out-performing traditional methods for dimensionality reduction in non-linear datasets. More recently, autoencoders have been explored as a replacement for PCA in various financial applications, encoding latent factors which account for non-linearities in return dynamics and allow for conditioning on exogenous covariates. \cite{AutoencoderAssetPricing}, for instance, use a series of encoders to estimate both latent factors and factor loadings (i.e. betas) used in conditional asset pricing models, demonstrating better out-of-sample pricing performance compared to traditional linear methods. \cite{AutoencoderTermStructure} explores the use of autoencoders to capture a low-dimensional representation of the term structure of commodity futures, and also shows superior reconstruction performance versus PCA. Given the strong performance enhancements of autoencoder architectures on non-linear datasets, replacing traditional factor models with autoencoders could be promising for co-movement measurement applications as well.

 In this article, we introduce the Autoencoder Reconstruction Ratio (ARR) -- a novel indicator to detect co-movement changes in real-time -- based on the average reconstruction error obtained from applying autoencoders to high-frequency returns data. Adopting the factor modelling approach, the ARR uses deep sparse denoising autoencoders to project second-level intraday returns onto a lower-dimension set of latent variables, and aggregates reconstruction errors up to the desired frequency (e.g. daily). Increased co-movement hence corresponds to periods where reconstruction error is low, i.e. when returns are largely accounted for by the autoencoder's latent factors. In line with the canonical Absorption Ratio, we evaluate the performance of the ARR by considering an application in systemic risk -- using increased co-movement of sector index returns to improve volatility and drawdown predictions for the total market.  The information content of the ARR is evaluated by measuring the performance improvements of machine learning predictors when the ARR is included as a covariate. Based on experiments performed for 4 prediction horizons (i.e. 5-min, 1-hour, 1-day, 1-week), the ARR was observed to significantly improve performance for volatility and market crashes forecasts over short-term (5-min and 1-hour) horizons.

\section*{Problem Definition}
For a portfolio of $N$ assets, let $r(t, n)$ be the returns of the $n$-th asset at time $t$ defined as below:
\begin{equation}
r(t,n) = \log p(t,n) - \log p(t-\Delta t, n),
\label{eqn:returns}
\end{equation}
where $p(t,n)$ is the value or price of asset $n$ at time $t$, and $\Delta t$ is a discrete interval corresponding to the desired sampling frequency. 

In their most general form, statistical factor models map a common set of $K$ latent variables to the returns of each asset as below:
\begin{align}
    r(t,n) &= \Bar{r}(t, n) + \epsilon(t, n) \label{eqn:factor_model}\\
    &= f_n\big(\bm{Z}(t)\big) + \epsilon(t, n), 
\end{align}
where the residual $\epsilon(t, n) \sim N\left(0, \xi_n^2\right)$ is the idiosyncratic risk of a given asset, and $f_n(.)$ is a generic function, and $\bm{Z}(t) = [z(t,1), \dots z(t, K)]^T$ is a vector of common latent factors.

Traditional factor models typically adopt a simple linear form, using asset-specific coefficients as below:
\begin{align}
    f_n\big(\bm{Z}(t)\big) =  \bm{\beta}(n)^T \bm{Z}(t),
\end{align}
where $\bm{\beta}(n)= [\beta(n,1), \dots \beta(n, K)]^T$ is a vector of factor loadings. While various approaches are available for latent variable estimation -- such as independent component analysis (ICA) \citep{ICAExample} or PCA \citep{CorrelationSpikesInBearMarketsMIT}  -- the number of latent variables are typically kept low to reduce the dimensionality of the dataset  (i.e. $K \ll N$).
\subsection*{Absorption Ratio}
A popular approach to measuring asset co-movement is the Absorption Ratio  \citep{AbsorptionRatio}, which performs dimensionality reduction on returns using PCA. Co-movement changes are then measured based on the total variance absorbed by a finite set of eigenvectors. With $K$ set to be a fifth of the number $N$ of available assets per \cite{AbsorptionRatio}, the Absorption ratio is defined as:
\begin{equation}
    \text{AR}(t) = \frac{\sum_{k=1}^K \sigma_{E_k}^2}{\sum_{n=1}^N \sigma_{A_n}^2},
\end{equation}
where $\text{AR}(t)$ is the Absorption Ratio at time $t$. $\sigma_{E_k}^2$ is the variance of the $k$-th largest eigenvector of the PCA decomposition, and $\sigma_{A_n}^2$ is the variance of the $n$-th asset.

\section*{Autoencoder Reconstruction Ratio}
To detect changes in asset co-movements, we propose the Autoencoder Reconstruction Ratio below, which can be interpreted as the normalised reconstruction mean squared error (MSE) of high-frequency returns within a given time interval:
\begin{align}
    \text{ARR}(t, \Delta t) =  \frac{\sum_{\tau=t-\Delta t}^{t} \sum_{n=1}^{N} \big( r(\tau,n) - \Bar{r}(\tau, n) \big)^2  }{\sum_{\tau=t-\Delta t}^{t} \sum_{n=1}^{N} r(\tau,n)^2 },
    \label{eqn:arr}
\end{align}
where $\Delta t$ is the time interval matching the desired sampling frequency, and $\Bar{r}(\tau, n)$ is the return of asset $n$ reconstructed by a deep sparse denoising autoencoder (see sections below). For our experiments, we train the autoencoder using one-second returns and compute ARRs across four different sampling frequencies (i.e. 5-min, 1-hour, 1-day, 1-week).

\subsection*{Relationship to the Absorption Ratio}
The ARR can also be interpreted as a slight reformulation of the standard Absorption Ratio, and we examine the relationship between the two in this section. 

Reintroducing the linear Gaussian assumptions used by the Absorption Ratio, we note that the denominator of Equation \eqref{eqn:arr} contains a simple estimator for the realised variance (RV) of each asset \citep{RealisedVariance}, i.e.:
\begin{align}
\text{RV}(t, \Delta t, n) &= \sum_{\tau=t-\Delta t}^t r(t,n)^2 \label{eqn:realised_vol}\\
& \approx \sigma_{A_n}^2.
\end{align}
 Moreover, by comparing to Equation \eqref{eqn:factor_model}, we can see that the numerator can be interpreted as an estimate of the sum of residual variances, leading to the form of the ARR below:
\begin{equation}
    \text{ARR}(t, \Delta t) = \frac{\sum_{n=1}^N \xi_n^2}{ \sum_{n=1}^N \sigma_{A_n}^2}.
\end{equation}
Assuming that residuals are uncorrelated between assets, we note that the sum of residual variances essentially corresponds to the variance unexplained by the selected factors. The ARR can in this case be expressed as:
\begin{align}
    \text{ARR}(t, \Delta t) &= \frac{\sum_{n=1}^N \sigma_{A_n}^2 -  \sum_{k=1}^K \sigma_{E_k}^2}{\sum_{n=1}^N \sigma_{A_n}^2} \\
    &= 1 - \text{AR}(t)
\end{align}
\subsection*{Autoencoder Architecture}
We adopt a deep sparse denoising autoencoder architecture \citep{DeepLearningBook} to compute reconstruction errors for the ARR. The network is predominantly divided into two parts -- 1) a decoder which reconstructs the returns vector from a reduced set of latent factors, and 2) an encoder which performs the low-dimensional projection -- both of which are described below.\\

\textbf{Decoder:}
\begin{align}
    \Bar{\bm{r}}(t) &= \bm{W}_{1}~\bm{h}_1(t) + \bm{b}_{1} \\
    \bm{h}_1(t) &= \text{ELU}(\bm{W}_{2}~\bm{Z}(t) + \bm{b}_{2})
\end{align}
where $\Bar{\bm{r}}(t) = [\Bar{r}(t,1), \dots, \Bar{r}(t,N)]^T$ is the vector of reconstructed returns, $\text{ELU}(.)$ is the exponential linear unit activation function \citep{elu}, $\bm{h}_1(t) \in \mathbb{R}^H$ is the hidden state of the decoder network, and $\bm{W}_1 \in \mathbb{R}^{N\times H}, \bm{W}_2 \in \mathbb{R}^{H \times K},$ $\bm{b}_1 \in \mathbb{R}^{N}, \bm{b}_2 \in \mathbb{R}^{H}$ are its weights and biases.\\

\textbf{Encoder:}
\begin{align}
    \bm{Z}(t) &= \text{ELU}\left(\bm{W}_{3}~\bm{h}_2(t) + \bm{b}_{3}\right) \\
    \bm{h}_2(t) &= \text{ELU}(\bm{W}_{4}~\bm{x}(t) + \bm{b}_{4}) 
\end{align}
where $\bm{Z}(t) \in \mathbb{R}^K$ is a low-dimensional projection of inputs $\bm{x}(t) = [r(t,1), \dots, r(t,N), \mathcal{T}]^T$, $\bm{h}_2(t) \in \mathbb{R}^H$ is the hidden state of the encoder network, and $\bm{W}_3 \in \mathbb{R}^{K\times H}, \bm{W}_4 \in \mathbb{R}^{H \times N},$ $\bm{b}_3 \in \mathbb{R}^{K}, \bm{b}_4 \in \mathbb{R}^{H}$ are its weights and biases. We note that the time-of-day $\mathcal{T}$ -- recorded as the number of seconds from midnight -- is also included in the input along with sector returns, allowing the autoencoder to account for any intraday seasonality present in the dataset.

To ensure that dimensionality is gradually reduced, we set $K = \left\lfloor N/5 \right\rfloor$ as per the original Absorption Ratio paper, and fix $H= \left\lfloor (N+K)/2 \right\rfloor$.

\subsection*{Network Training}
To improve generalisation on test data, sparse denoising autoencoders introduce varying degrees of regularisation. Firstly, a sparsity penalty in the form of a L1 regularisation term is added to the reconstruction loss as below:
\begin{align}
    \mathcal{L}(\bm{\Theta}) 
    =& \frac{\sum_{t=1}^T \sum_{n=1}^N (r(t,n)  - \Bar{r(t,n)})^2}{TN} \nonumber\\
    &+ \alpha \lVert \bm{Z}(t) \rVert_1,
\end{align}
where $\bm{\Theta}$ represents all network weights,  $\alpha$ is a penalty weight which we treat as a hyperparameter, and $\lVert . \rVert_1$ is the L1 norm. Secondly, inputs are corrupted with noise during training -- forcing the autoencoder to learn more general relationships from the data. Specifically, we adopt masking noise for our network, which simply corresponds to the application of dropout \citep{dropout} to encoder inputs.

Hyperparameter optimisation is performed using 20 iterations of random search, with networks trained up to a maximum of 100 epochs per search iteration. Full hyperparameter details are listed in the appendix for reference.

\section*{Forecasting Systemic Risk with the ARR}
To demonstrate the utility of the ARR, we consider applications in systemic risk forecasting -- using ARRs computed from subsector indices to predict risk metrics associated with the overall market. Specifically, we consider the two key use-cases for the ARR:
\begin{enumerate}
    \item \textbf{Volatility Forecasting} -- i.e. predicting market turbulence as measured by spikes in realised volatility;
    \item \textbf{Predicting Market Crashes} -- i.e. providing an early warning signal for sudden market declines, allowing for timely risk management.
\end{enumerate}

\subsection*{Description of Dataset}
We focus on the total US equity market and 11 constituent sector indices for our investigation, using high frequency total returns sampled every second to compute the ARR. Intraday index data from the Center of Research in Security Prices (CRSP) was downloaded via Wharton Research Data Services (\cite{wrds}) from 2012-12-07 to 2019-03-29, with the full list of indices provided in Exhibit \ref{tab:index_data}.

\begin{table}[h]
\begin{tabular}{@{}ll@{}}
\toprule
\textbf{Ticker}  & \textbf{Index Description}                         \\ \midrule
\textbf{CRSPTMT} & CRSP US Total Market Total-Return Index      \\ \midrule
\textbf{CRSPRET} & CRSP US REIT Total-Return Index              \\
\textbf{CRSPENT} & CRSP US Oil and Gas Total-Return Index       \\
\textbf{CRSPMTT} & CRSP US Materials Total-Return Index         \\
\textbf{CRSPIDT} & CRSP US Industrials Total-Return Index       \\
\textbf{CRSPCGT} & CRSP US Consumer Goods Total-Return Index    \\
\textbf{CRSPHCT} & CRSP US Health Care Total-Return Index       \\
\textbf{CRSPCST} & CRSP US Consumer Services Total-Return Index \\
\textbf{CRSPTET} & CRSP US Telecom Total-Return Index           \\
\textbf{CRSPUTT} & CRSP US Utilities Total-Return Index         \\
\textbf{CRSPFNT} & CRSP US Financials Total-Return Index        \\
\textbf{CRSPITT} & CRSP US Technology Total-Return Index        \\ \bottomrule
\end{tabular}
\caption{US Maket \& Sector Indices Used in Tests}
\label{tab:index_data}
\end{table}

\subsection*{Reconstruction Performance}
We compare the reconstruction accuracy of the deep sparse denoising autoencoder and standard PCA to evaluate the benefits of a non-linear approach to dimensionality reduction. To train the autoencoder, we divide the data into a training set used for network backpropagation (2012-2014), a validation set used for hyperparameter optimisation (2015 only), and a out-of-sample test set (2016-2019) used to quantify the information content of the ARR. PCA factors were calibrated using covariance matrix estimated with data from 2012-2016, keeping the same out-of-sample data as the autoencoder. The returns vector is projected onto 2 latent variables, with the dimensionality of the latent space taken to be $1/5$ the number of indices used -- as per the original Absorption Ratio paper \citep{AbsorptionRatio}.

We evaluate the reconstruction accuracy using the test dataset, based on the R-squared ($R^2$) of reconstructed returns for each prediction model. Given that the ARR is computed based on sector data alone, we only focus on the reconstruction accuracy of the 11 sector indices. We also test for the significance of the results with a bootstrap hypothesis test -- using 500 bootstrap samples and adopting the null hypothesis that there is no difference in $R^2$ between the autoencoder and PCA reconstruction.

\begin{table}[h]
\centering
\begin{tabular}{@{}l|cc|c@{}}
\toprule
{\ul \textbf{}} & \textbf{PCA} & \textbf{Autoencoder} & \textbf{P-Value} \\ \midrule
$\bm{R^2}$                     & 0.340    & \textbf{0.461*}    &  $<0.01$   \\ \bottomrule
\end{tabular}
\caption{Out-of-Sample Reconstruction $R^2$ (2016-2019)}
\label{tab:reconstruction_results}
\end{table}
From the results in Exhibit \ref{tab:reconstruction_results}, we can see that the autoencoder greatly enhances reconstruction accuracy -- increasing the out-of-sample $R^2$ by more than $35\%$ and statistically significant with $99\%$ confidence. These improvements highlight the benefits of adopting a non-linear approach to dimensionality reduction, and the suitability of the autoencoder for modelling intraday returns. Such an approach can have wider applications in various areas of quantitative finance.

\subsection*{Empirical Analysis}
Next, we perform an exploratory investigation into the relationships between the total market index, and the ARR of its constituent sectors. Specifically, we examine the following metrics aggregated over different sampling frequencies ($\Delta t \in \{ 5 \text{-min}, 1 \text{-hour}, 1 \text{-day}, 1 \text{-week} \})$:
\begin{itemize}
    \item \textbf{Returns} -- Corresponding to the difference in log prices based on Equation \eqref{eqn:returns}.
    \item \textbf{Log Realised Volatility (Log RV)} -- Computed based on the logarithm of the simple realised volatility estimator of Equation \eqref{eqn:realised_vol}.
    \item \textbf{Drawdowns (DD)} -- Determined by the factional decrease from the maximum index value from the start of our evaluation period.
\end{itemize}

\begin{figure*}[ptbh]
\centering
\begin{subfigure}[]{0.24\linewidth}
\includegraphics[width=1\linewidth]{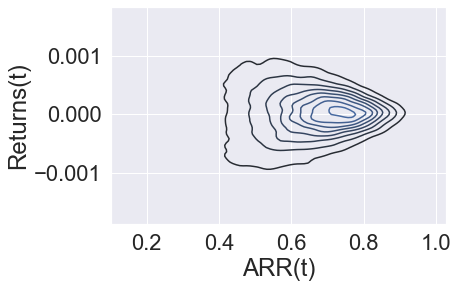}
\caption{$\Delta t = 5 \text{-min}$}
\end{subfigure}
\begin{subfigure}[]{0.24\linewidth}
\includegraphics[width=1\linewidth]{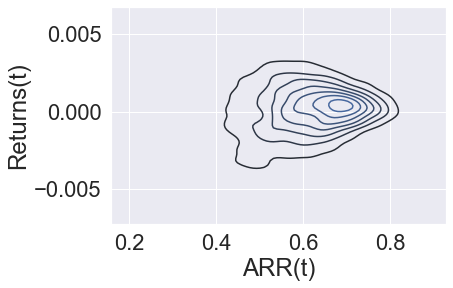}
\caption{$\Delta t = 1 \text { hour}$}
\end{subfigure}
\begin{subfigure}[]{0.24\linewidth}
\includegraphics[width=1\linewidth]{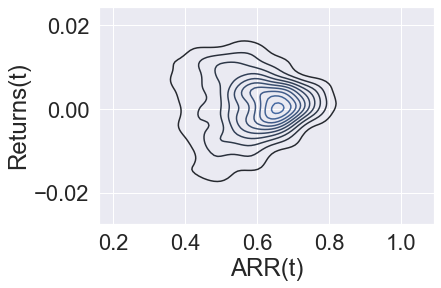}
\caption{$\Delta t = 1 \text{-day}$}
\end{subfigure}
\begin{subfigure}[]{0.24\linewidth}
\includegraphics[width=1\linewidth]{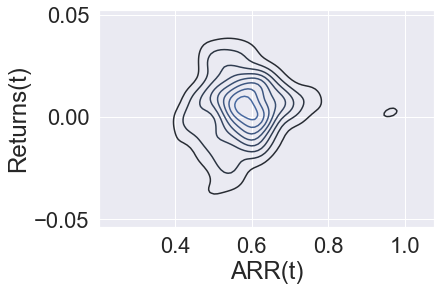}
\caption{$\Delta t = 1 \text{-week}$}
\end{subfigure}
\caption{In-sample  KDE Plots for Returns vs. ARR.}
\label{fig:kde_returns_aar}
\end{figure*}

\begin{figure*}[ptbh]
\centering
\begin{subfigure}[]{0.24\linewidth}
\includegraphics[width=1\linewidth]{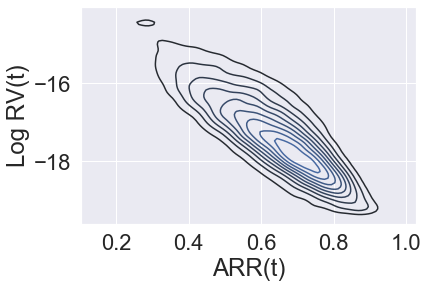}
\caption{$\Delta t = 5 \text{-min}$}
\end{subfigure}
\begin{subfigure}[]{0.24\linewidth}
\includegraphics[width=1\linewidth]{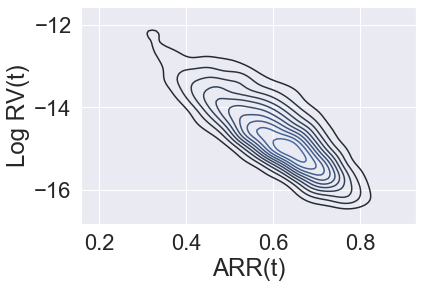}
\caption{$\Delta t = 1 \text { hour}$}
\end{subfigure}
\begin{subfigure}[]{0.24\linewidth}
\includegraphics[width=1\linewidth]{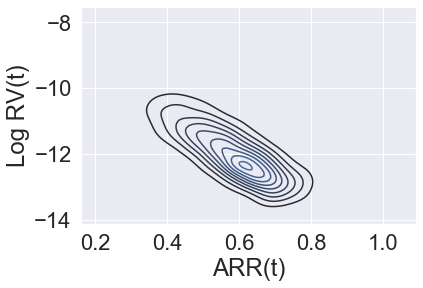}
\caption{$\Delta t = 1 \text{-day}$}
\end{subfigure}
\begin{subfigure}[]{0.24\linewidth}
\includegraphics[width=1\linewidth]{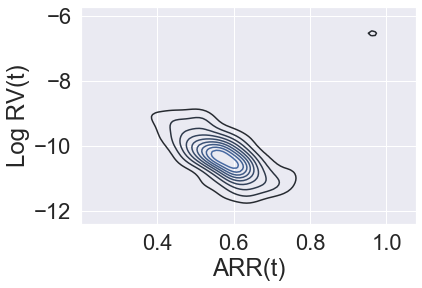}
\caption{$\Delta t = 1 \text{-week}$}
\end{subfigure}
\caption{In-Sample KDE Plots for Log RV vs. ARR.}
\label{fig:kde_rv_aar}
\end{figure*}

\begin{figure*}[ptbh]
\centering
\begin{subfigure}[]{0.24\linewidth}
\includegraphics[width=1\linewidth]{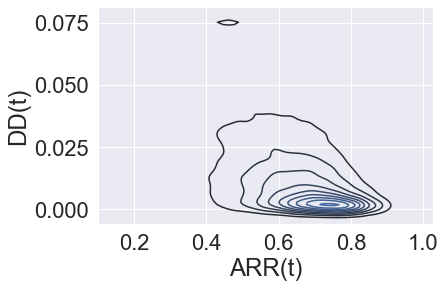}
\caption{$\Delta t = 5 \text{-min}$}
\end{subfigure}
\begin{subfigure}[]{0.24\linewidth}
\includegraphics[width=1\linewidth]{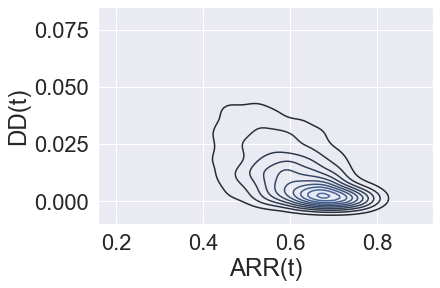}
\caption{$\Delta t = 1 \text { hour}$}
\end{subfigure}
\begin{subfigure}[]{0.24\linewidth}
\includegraphics[width=1\linewidth]{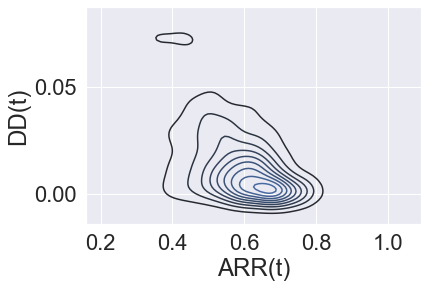}
\caption{$\Delta t = 1 \text{-day}$}
\end{subfigure}
\begin{subfigure}[]{0.24\linewidth}
\includegraphics[width=1\linewidth]{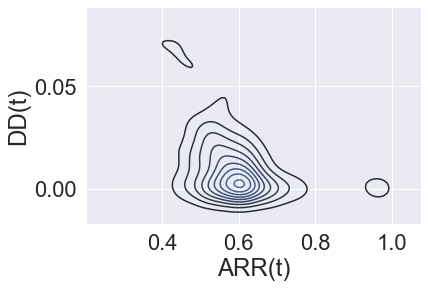}
\caption{$\Delta t = 1 \text{-week}$}
\end{subfigure}
\caption{In-Sample KDE Plots for Drawdowns vs. ARR.}
\label{fig:kde_mdd_aar}
\end{figure*}

ARRs were similarly computed for the 4 sampling frequencies, and aggregated based on Equation \eqref{eqn:arr}. Using the KDE plots of Exhibits \ref{fig:kde_returns_aar} to \ref{fig:kde_mdd_aar}, we perform an in-sample (2012-2015) empirical analysis of the coincident relationships between the ARR and the metrics above. To avoid spurious effect, we remove the eves of public holidays from our analysis -- where markets are open only in the morning and trading volumes are abnormally low. In addition, we winsorise the data at the $1^{st}$ and $99^{th}$ percentiles, reducing the impact of outliers to improve data visualisation.
\begin{figure*}[ptbh]
\centering
\includegraphics[width=0.97\linewidth]{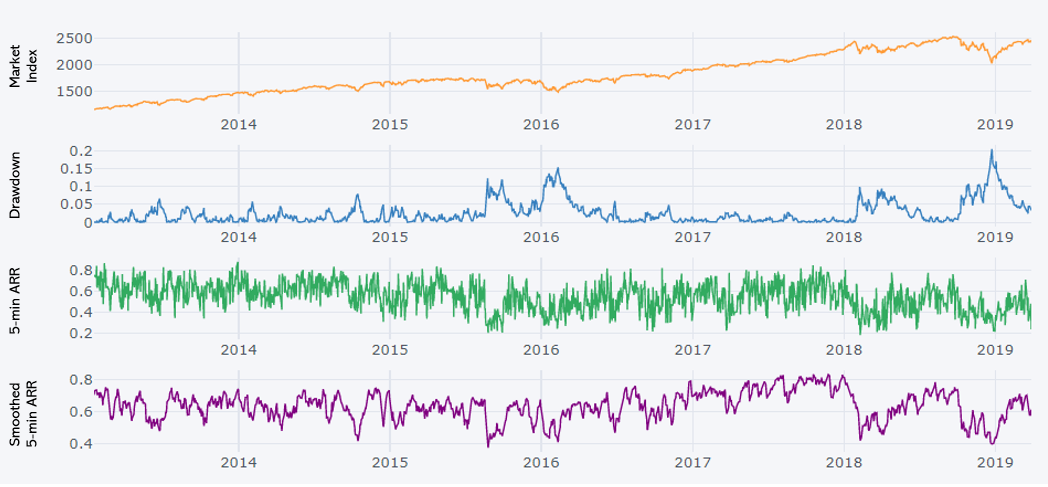}
\caption{5-min ARR from 2012 to 2019.}
\label{fig:arr_evolution}
\end{figure*}

Looking at the KDE plots for returns against the ARR in Exhibit \ref{fig:kde_returns_aar}, we observe a noticeable increase in the dispersion of returns appears at low ARR values. This is echoed by the KDE plots for Log RV in \ref{fig:kde_rv_aar} -- which appears to increase linearly with decreasing ARR. Drawdowns behave in a similar fashion as well, with larger drawdowns appearing to occur at low values of the ARR. Effects also are consistent across different horizons, with similar patterns observed for all sampling frequencies. On the whole, the results validate the findings observed in previous works -- indicating that increased asset co-movements do indeed coincide with periods of market weakness -- with lower ARR values observed around periods of high volatility or high drawdowns.

To visualise how the ARR changes over time, we plot 5-min ARR (i.e. ARR$(t, 5 \text{-min})$) against the prices and drawdowns of the CRSP Total Market Index in Exhibit \ref{fig:arr_evolution}. Given the noisiness of the raw ARR values, we also included a smoothed version of 5-min ARR in the bottom -- computed based on an exponentially weighted moving average with a 1-day half-life. We can see from the results that dips in the ARR occur slightly before large drawdown periods, particularly around the sell-offs of August 2015 and early 2018. As such, the ARR could potentially be used to improve predictions of volatility spikes or sudden market crashes in the near future -- which we further investigate in the next section.

\section*{Quantifying the Information Content \\of the ARR}
We attempt to quantify the information content of the ARR by observing how much it improves forecasting models for the risk metrics above. To do so, we adopt several machine learning benchmarks for risk prediction, and evaluate the models' forecasting performance both with and without the ARR included in its inputs. This allows us to evaluate how much additional information is provided by the ARR, above that provided by temporal evolution of realised volatility or drawdowns alone.

\subsection*{Benchmark Models}
Given the non-linear relationships observed between returns and drawdowns versus the ARR, we adopt a series of machine learning benchmarks on top of linear models. Specifically, we consider 1) linear/logistic regression, 2) gradient boosted decision trees (GBDTs), and 3) simple multi-layered perceptrons (MLPs) for our forecasting applications. Hyperparameter optimisation is performed using up to 200 iterations of random search, with optimal hyperparameters selected using 3-fold cross validation. Additional training details can also be found in the appendix.

\subsection*{Volatility Forecasting Methodology}
We treat volatility forecasting as a regression problem, focusing on predicting Log RV over different horizons -- using a variant of the HAR-RV model of \cite{HARRV}. For 5-min Log RV, a simple HAR-RV model can take the form below, incorporating log RV terms for longer horizons into the forecast:
\begin{align}
    &\Bar{\nu}(t+5\text{-min}, 5\text{-min})&& \nonumber\\
    &=  \beta_1~\nu(t, 5 \text{-min}) + \beta_2~\nu(t, 1\text{-hour})&& \nonumber\\
    &+ \beta_3~\nu(t, 1\text{-day}) + \beta_4~\nu(t, 1\text{-week}),&&
\end{align}

where $\Bar{\nu}(t+5\text{-min}, 5\text{-min})$ is 5-min log RV predicted over the next time-step, $\nu(t, \Delta t)$ is the log RV from $t-\Delta t$ to $t$, and $\beta_{(.)}$ are linear coefficients. 

We adopt a similar non-linear variant for our forecast for each prediction horizon $\delta$. For tests with the ARR, this takes the form:
\begin{align}
    \Bar{\nu}(t+\delta, \delta) = g_1(\psi(t), \omega(t)), 
    \label{eqn:vol_forecasting}
\end{align}

where $\delta \in \Psi$, $\Psi = \{ 5 \text{-min}, 1 \text{-hour}, 1 \text{-day}, 1 \text{-week}  \}$, $\psi(t) = \{ \nu(t, d): d \in \Psi \text{ and } d>= \delta \}$, $\omega(t) = \{ \text{ARR}(t, d): d \in \Psi \text{ and } d>= \delta \}$, and $g_2(.)$ is a prediction model mapping inputs to returns. The combination of both realised volatility and ARR values hence allows us to determine if the ARR supplies any information above that provide the volatility time series alone.

For tests without the ARR, we continue to use Equation \eqref{eqn:vol_forecasting} but omit ARR values, i.e. $\omega(t) = \emptyset$. We evaluate regression performance using the $R^2$ of forecasted log RV.

\begin{table*}[tbh]
\centering
\begin{subtable}[b]{0.45\textwidth}
\centering
\begin{tabular}{@{}ll|llll@{}}
\toprule
\textbf{}       &          & \textbf{5-min}  & \textbf{1-hour}    & \textbf{1-day}     & \textbf{1-week}    \\ \midrule
\textbf{Linear} & With ARR & \textbf{0.635*}           & 0.496              & 0.389              & 0.579              \\
\textbf{}       & No ARR   & 0.626           & \textbf{0.536*}              & \textbf{0.431*}              & \textbf{0.611*}              \\
\textbf{}       & P-Values & \textless{}0.01 & \textgreater{}0.99 & \textgreater{}0.99 & \textgreater{}0.99 \\ \midrule
\textbf{GBDT}   & With ARR & \textbf{0.635*}           & \textbf{0.554*}              & 0.425              & 0.399              \\
\textbf{}       & No ARR   & 0.627           & 0.536              & \textbf{0.434}              & \textbf{0.593*}              \\
\textbf{}       & P-Values & \textless{}0.01 & \textless{}0.01    & 0.821              & \textgreater{}0.99 \\ \midrule
\textbf{MLP}    & With ARR & \textbf{0.641*}           & \textbf{0.571*}              & 0.387              & 0.527              \\
\textbf{}       & No ARR   & 0.631           & 0.549              & \textbf{0.426*}              & \textbf{0.636*}              \\
\textbf{}       & P-Values & \textless{}0.01 & \textless{}0.01    & \textgreater{}0.99 & \textgreater{}0.99 \\ \bottomrule
\end{tabular}
\caption{$R^2$ for Log RV Predictions}
\label{tab:log_rv_r2}
\end{subtable}\hfill
\begin{subtable}[b]{0.45\textwidth}
\centering
\begin{tabular}{@{}ll|llll@{}}
\toprule
\textbf{}         &          & \textbf{5-min}  & \textbf{1-hour} & \textbf{1-day} & \textbf{1-week} \\ \midrule
\textbf{Linear} & With ARR & \textbf{0.598*} & \textbf{0.629*} & \textbf{0.585} & \textbf{0.372}  \\
\textbf{}         & No ARR   & 0.587           & 0.570           & 0.582          & 0.333           \\
\textbf{}         & P-Values & \textless{}0.01 & \textless{}0.01 & 0.464          & 0.284           \\ \midrule
\textbf{GBDT}     & With ARR & \textbf{0.590*} & \textbf{0.562*} & \textbf{0.529} & \textbf{0.605}  \\
\textbf{}         & No ARR   & 0.567           & 0.516           & 0.465          & 0.575           \\
\textbf{}         & P-Values & \textless{}0.01 & 0.01           & 0.138          & 0.359           \\ \midrule
\textbf{MLP}      & With ARR & \textbf{0.589}  & \textbf{0.564}  & \textbf{0.586} & \textbf{0.423}  \\
\textbf{}         & No ARR   & 0.588           & 0.558           & 0.540          & 0.210           \\
\textbf{}         & P-Values & 0.374           & 0.392           & 0.234          & \textless{}0.01 \\ \bottomrule
\end{tabular}
\caption{AUROC for Crash Predictions}
\label{tab:crash_auroc}
\end{subtable}
\caption{Out-of-Sample Forecasting Results With and Without ARR Inputs.}
\label{tab:forecasting_results}
\end{table*}

\subsection*{Crash Prediction Methodology}
We take a binary classification approach to forecasting market crashes, using our benchmark models to predict the onset of a sharp drawdown. First, we define a $z$-score metric for returns $\gamma(t, \lambda)$ as below:
\begin{equation}
    \gamma(t, \lambda) = \frac{r(t) - m(t, \lambda)}{s(t, \lambda)},
\end{equation}
where $r(t)$ is the return for the total market index, $m(t, \lambda)$ is the exponentially weighted moving average for returns with a half-life of $\lambda$, and $s(t, \lambda)$ its exponentially weighted moving standard deviation.

Next, based on our z-score metric, we define a market crash to be a sharp decline in market returns, i.e.:
\begin{align}
    c(t) = \mathbb{I}(\gamma(t,\lambda) < C),
\end{align}
where $c(t) \in {1, 0}$ is a crash indicator, $\mathbb{I}(.)$ is an indicator function, and $C$ is the $z$-score threshold for returns. For our experiments, we set $\lambda$ to be 10 discrete time steps and $C=-1.5$.

We then model crash probabilities using a similar form to Equation \eqref{eqn:vol_forecasting}:
\begin{align}
    p(c(t) = 1) = g_2(\psi(t), \omega(t)), 
\end{align}
where $g_2(.)$ is a function mapping inputs to crash probabilities $p(c(t) = 1)$.

Given that crashes are rare by definition -- with $c(t)=1$ for less than $10\%$ of time steps for daily frequencies -- we also oversample the minority class to address the class imbalance problem.  Classification performance is evaluated using the area under the receiver operating characteristic (AUROC).

\subsection*{Results and Discussion}
The results for both volatility forecasting and market crash prediction can be found in Exhibit \ref{tab:forecasting_results}, both including and excluding ARR values. To determine the statistical significance of improvements, we conduct a bootstrap hypothesis test under the null hypothesis that performance results are better when the ARR is included, using a non-parametric bootstrap with 500 samples. 

From the volatility forecasting $R^2$ values in Exhibit \ref{tab:log_rv_r2}, we can see that the ARR consistently improves 5-min and 1-hour forecasts for all non-linear models, and significant improvements are observed for linear models for 5-min sampling frequencies. This indicates that the ARR is informative for short-term forecasts, and can help enhance risk predictions in the near-term. For longer horizons however (i.e. 1-day and 1-week) we observe that the inclusion of the ARR reduces prediction accuracy -- potentially indicating the presence of overfitting on the training set when ARRs are introduced.

For crash predictions AUROC results in Exhibit \ref{tab:crash_auroc}, we note that the ARRs are observed to improve forecasts for all models and sampling frequencies -- with statistical significance at the $99\%$ level observed for both linear and GBDT forecasts over shorter horizons. This echoes the volatility forecasting results -- indicating that ARRs can be useful to inform short-term risk predictions. 

\section*{Conclusions}
We introduce the Autoencoder Reconstruction Ratio (ARR) in this paper, using it as a real-time measure of asset co-movement. The ARR is based on the normalised reconstruction error of a deep sparse denoising autoencoder applied to a basket of asset returns -- which condenses the returns vector onto a lower dimensional set of latent variables. This replaces the PCA modelling approach used by the Absorption Ratio of \cite{AbsorptionRatio}, which allows the ARR to better model returns that violate basic PCA assumptions (e.g. non-Gaussian returns). Through experiments on a basket of 11 CRSP US sector indices, we demonstrate that the autoencoder significantly improves the out-of-sample reconstruction performance when compared to PCA, increasing the combined $R^2$ by more than $35\%$.

Given the links identified between increased asset co-movements and the fragility of the overall market in previous works \citep{AbsorptionRatio,IncreasedCorrelationInBearMarkets}, we also evaluate the use of the ARR in a systemic risk application. First, we conduct an empirical analysis of the relationship between risk metrics of the combined market (using the CRSP US Total Market Index as a proxy) and the ARR computed from its sub sectors of the market. Based on an analysis of the KDE plots of risk metrics vs. ARRs, we show that low values of the ARR coincide with high volatility and high drawdown periods in line with previous findings. Next, we evaluate the information content of the ARR by testing how much the ARR improves risk predictions for a various benchmark models. We find that the ARR is informative for both volatility and market crash predictions over short horizons, and significantly increases 5-min and 1-hour forecasting performance across most model benchmarks.

{\footnotesize
\bibliographystyle{authordate}
\bibliography{aar_bib}
}
\appendix
\subsection{Additional Training Details}
\subsubsection*{Python Libraries} Deep sparse denoising autoencoders are defined and trained using the \texttt{TensorFlow}  \citep{tensorflow2015-whitepaper}. For Gradient Boosted Decision Trees, we use the \texttt{LightGBM} library \citep{lightgbm} -- using the standard \texttt{LightGBMRegressor} and \texttt{LightGBMClassifier} depending on the forecasting problem. The remainder of the models are implemented using standard \texttt{scikit-learn} classes \citep{scikit-learn} -- with classes described in the hyperparameter optimisation section.\\

\subsubsection*{Hyperparameter Optimisation Details} Random search is conducted by sampling over a discrete set of values for each hyperparameter, which are listed below for each hyperparameter. For ease of reference, hyperparameters for all \texttt{scikit-learn} and \texttt{LightGBM} classes are referred to by the default argument names used in their respective libraries.\\

\textbf{Deep Sparse Denoising Autoencoder}
\begin{itemize}
    \item Dropout Rate -- [0.0, 0.2, 0.4, 0.6, 0.8]
    \item Regularisation Weight $\alpha$ -- [0.0, 0.01, 0.1, 1.0, 10]
    \item Minibatch Size -- [256, 512, 1024, 2048]
    \item Learning Rate -- [$10^{-5}$, $10^{-4}$, $10^{-3}$, $10^{-2}$, $10^{-1}$, 1.0]
    \item Max. Gradient Norm -- [$10^{-4}$, $10^{-3}$, $10^{-2}$, $10^{-1}$, 1.0, 10.0]
\end{itemize}~\\

\textbf{Linear Regression}
\begin{itemize}
    \item Package Name -- \texttt{sklearn}.\texttt{linear\_model}
    \item Class Name -- \texttt{LogisticRegression}
    \item 'alpha' -- [$10^{-5}$, $10^{-4}$, $10^{-3}$, $10^{-2}$, $10^{-1}$, 1, 10, $10^{2}$],
    \item 'fit\_intercept' -- [False, True],
\end{itemize}~\\

\textbf{Logistic Regression}
\begin{itemize}
    \item Package Name -- \texttt{sklearn}.\texttt{linear\_model}
    \item Class Name -- \texttt{LogisticRegression}
    \item 'penalty' -- ['l1']
    \item 'C' -- [0.01, 0.1, 1.0, 10, 100]
    \item 'fit\_intercept' -- [False]
    \item 'solver' --['liblinear']
\end{itemize}~\\

\textbf{Gradient Boosted Decision Tree}
\begin{itemize}
    \item Package Name -- \texttt{lightgbm}
    \item Class Name -- \texttt{LGBMRegressor} or \texttt{LGBMClassifier}
    \item 'learning\_rate' -- [$10^{-4}$, $10^{-3}$, $10^{-2}$, $10^{-1}$]
    \item 'n\_estimators' -- [5, 10, 20 ,40, 80, 160, 320]
    \item 'num\_leaves' -- [5, 10, 20, 40, 80]
    \item 'n\_jobs' -- [5]
    \item 'reg\_alpha' -- [0, $10^{-4}$, $10^{-3}$, $10^{-2}$, $10^{-1}$]
    \item 'reg\_beta' -- [0, $10^{-4}$, $10^{-3}$, $10^{-2}$, $10^{-1}$]
    \item 'boosting\_type' -- ['gbdt']
\end{itemize}~\\

\textbf{Multi-layer Perceptron}
\begin{itemize}
    \item Package Name -- \texttt{sklearn}.\texttt{neural\_network}
    \item Class Name -- \texttt{MLPRegressor} or \texttt{MLPClassifier}
    \item 'hidden\_layer\_sizes' -- [5, 10, 20, 40, 80, 160]
    \item 'activation' -- ['relu']
    \item 'alpha' -- [0, $10^{-4}$, $10^{-3}$, $10^{-2}$, $10^{-1}$, 1, 10, $10^{2}$]
    \item 'learning\_rate\_init' -- [$10^{-4}$, $10^{-3}$, $10^{-2}$, $10^{-1}$]
    \item 'early\_stopping' -- [True],
    \item 'max\_iter' -- [500]
\end{itemize}

\end{document}